\documentclass[parskip=half-, fontsize = 10pt]{scrartcl} %10pt,
\usepackage[margin=1in]{geometry}
\usepackage{amsmath}
\usepackage{amssymb}
\usepackage{algorithm,algpseudocode}
\usepackage{algpseudocode}
\usepackage{nicefrac}
\usepackage{array}
\usepackage{dirtytalk}
\usepackage{placeins}
\usepackage{hyperref}

\DeclareMathOperator*{\arginf}{arg\,inf}
\usepackage{latexsym}
\usepackage{stmaryrd}
\usepackage{titling}
\usepackage{xcolor}
\usepackage{amsthm}
\definecolor{changed}{HTML}{000000}
\usepackage{float}
\usepackage[autostyle=true]{csquotes}
\usepackage{mathtools}
\theoremstyle{definition}

\usepackage{caption}
\newcommand{\PWM}{\textcolor{changed}{ideal-world-model} }
\newcommand{\LWM}{\textcolor{changed}{internal-world-model} }
\title{\textbf{Outsourcing Control requires Control Complexity}} %an increased Controller Complexity
\author{Carlotta Langer$^{1 }$ and Nihat Ay$^{1, 2, 3}$
\\
\normalsize{$^{1}$ \, Hamburg University of Technology, Hamburg, Germany }\\
\normalsize{$^{2}$ \, Santa Fe Institute, Santa Fe, USA}\\
\normalsize{$^{3}$ \, Leipzig University, Leipzig, Germany}} 
\date{}
\usepackage[backend=bibtex ,natbib=true,giveninits=true,maxnames=25]{biblatex}
\usepackage{arydshln}
\begin{document}
\maketitle
\begin{abstract}
%An embodied agent influences its environment and is being influenced by it. We use the sensorimotor loop to model the interactions among an agent’s brain, body and environment. Thereby we can calculate various information theoretic measures that quantify different information flows in the system, one of which corresponds to Integrated Information. The Integrated Information Theory provides a quantitative approach to consciousness and can be applied to neural networks. Additionally we are able to measure the interaction among the body and the environment, which leads to the concept of Morphological Computation. Previous research reveals an antagonistic relationship between Integrated Information and Morphological Computation. A morphology adapted well to a task can reduce the necessity for Integrated Information significantly. This creates the problem that embodied intelligence is correlated with reduced conscious experience. Here we propose a solution to this problem. More precisely, we theorize that a high Integrated Information value is necessary for an agent to predict the next sensory state. We demonstrate the dynamics of the measures in a simple experimental setup in which the agents learn by using the em-algorithm. The results support the hypothesis that Integrated Information is necessary for learning.

An embodied agent influences its environment and is influenced by it. We use the sensorimotor loop to model these interactions and quantify the information flows in the system by information theoretic measures. This includes a measure for the interaction among the agent's body and its environment, often referred to as Morphological Computation. Additionally, we examine the controller complexity, which can be seen in the context of the Integrated Information Theory of consciousness.
Applying this framework to an experimental setting with simulated agents allows us to analyze the interaction between an agent and its environment, as well as the complexity of its controller. 
Previous research reveals that a morphology adapted well to a task can reduce the required complexity of the controller \textcolor{changed}{substantially}. %, hence Morphological Computation and the controller complexity have a.. dynamic. 
%This creates the problem that embodied intelligence is correlated with a reduced necessity of a controller, a brain. 
In this work we observe that the agents first have to understand the relevant dynamics of the environment in order to interact well with their surroundings. Hence, an increased controller complexity can facilitate a better interaction between an agent's body and its environment.  

\end{abstract}
\section{Introduction}

Every embodied agent, whether it is an animal, a human or a robot, exists in constant interaction with its environment. The morphology of an agent's body has a significant impact on the nature of this interaction. The authors of the book \enquote{How the body shapes the way we think: a new view of intelligence} \citep{Pfeifer2006}  emphasize the importance of this interaction and the influence of it on the structure of the control architecture, i.e.~the brain of an agent. \citet{Pfeifer2009} express this notion more precisely in the following way:
\begin{quote}
``{[...] there is a kind of trade-off or balance: the better the exploitation of the dynamics,
the simpler the control, the less neural processing will be required.}''
\end{quote}

This suggests that the way an agent interacts with its environment has an impact on the complexity of its control architecture. Our previous research with simulated agents confirms this intuition \citep{Langer2021}. \textcolor{changed}{There we observe that a better interaction with the environment reduces the necessity for a complex controller.} This relationship insinuates that a sufficiently well-designed morphology might make a complex control architecture superfluous. 

In this work we extend the framework to include the process of learning a new task. Thereby we are able to observe that a high controller complexity can facilitate a better interaction with the environment, captured by a measure called ``Morphological Computation''. Furthermore, agents with a simplified control architecture seem to be almost unable to learn a good strategy. Hence, we conclude that agents need an increased controller complexity in order to learn and that both concepts, the controller complexity and Morphological Computation, influence each other. 

\textcolor{changed}{
In the next section we describe the historical background and outline, before we discuss the intuition in more detail.}

\subsection*{Historical Background and Outline}

In this work we analyze the dynamics of the information flows in simple, simulated agents. \textcolor{changed}{Here we apply methods from information theory, which is based on the mathematical theory of communication introduced by \cite{Shannon}. There the author quantifies the properties of a communication channel and we use related information theoretic measures to quantify information flows among an agent's body, controller and environment. These information flows are modeled by the sensorimotor loop, similar to the approaches applied in, for example, \citep{Polani2007, Tishby2011, Polani2009}.}

\textcolor{changed}{
The sensorimotor loop reflects the interactions among the elements of the sensors $S$, the actuators $A$ and the controller $C$ and can be translated to probability distributions that define the behavior of the agents. In our experiments the agents are faced with the task of moving through a racetrack environment} without touching the walls. This is discussed in more detail in the section \enquote{Setting of the Experiment}. Additionally, some agents, the \LWM agents, are also equipped with an internal prediction of the next sensory state $S'$  and we call the mechanism that generates this state the \emph{internal world model}. Such an internal world model was also used by \citet{Ay2013, Ay2013a}. This approach allows us to analyze the information flows among the different parts of the agents, and especially the prediction, in detail. An overview over the different agents can be found in Figure \ref{fig:3} in the section \enquote{The Agents and the World Models}.

The applied learning algorithm is defined in the section \enquote{Learning} and it is based on a modification of the em-algorithm, a well-known information geometric algorithm \citep{Csiszar1984, Amari1995}.
We combine two different instances of this algorithm in order to alternate between optimizing the agent's behavior and updating its world model. Optimizing the behavior is done by maximizing the likelihood of a goal variable and this follows the reasoning of the approach described by \citet{Attias}, and further analyzed by \citet{Toussaint2006, Toussaint2009}, called \emph{Planning as Inference}. We also use this optimization in \citep{Langer2021}. 

While the agents learn we calculate various information theoretic measures, defined in the section \enquote{Measures of the Information Flow}. \textcolor{changed}{One important aspect is to assess the complexity of the controller, which we determine using two different measures that quantify distinct mechanisms of the controller. There exist various approaches to complexity. In this work we consider a system to be complex if it is more than the sum of its parts. Hence, the first measure that contributes to the controller complexity quantifies how much information integration exists between two parts of the controller. If we are able to divide the controller into two distinct parts without loss of functionality, then we call it split and not complex.}  

This measure can be seen in the context of the Integrated Information Theory (IIT) of consciousness, originally proposed by Tononi. The core idea of IIT is that the level of consciousness of a system can be equated with the amount of information integration among different parts of it. This theory developed rapidly from a measure for brain complexity \citep{Tononi1994} towards a broad theory of consciousness \citep{Tononi2004, Oizumi2014, Barbosa2021}. 
Hence, there exist various types of Integrated Information measures depending on the version of the theory these measures are based on and the setting they are defined in. Here, we use the information geometric measure that we propose in \citep{Langer} as a measure for the controller complexity. Thereby, we follow the suggestion by \citet{Mediano} to adopt a more pragmatic point of view on Integrated Information measures.

\textcolor{changed}{Additionally, we calculate a measure of synergy of the internal world model in order to assess the controller complexity. This internal world model predicts the next sensory state and is vital for finding an optimal behavior. Here we measure the importance of the interplay between the different information flows going to the internal world model and we call this the \textit{synergistic prediction}}.

\textcolor{changed}{The term \textit{synergistic} suggests a relation between this measure and the context of the partial information decomposition of random variables. There the goal is to decompose the information that a set of variables holds about a target variable into separate, non-negative terms, namely into redundant, synergistic and unique information, introduced by \cite{Williams2010}. There exist different definitions of these terms, for instance, the BROJA partial information decomposition in the case of two input variables, defined in \citep{Bertschinger}. Using a similar approach to synergy as we apply here leads to a definition of unique information in Section 3.5 of \citep{Ghazi-Zahedi2019}. Alternatively, a measure for representational complexity of feed-forward networks is discussed by \cite{ehrlich2023a}. This quantifies how much of a system needs to be observed simultaneously to access a particular piece of information.}

In \citep{Langer2021} we compare the \textcolor{changed}{controller complexity  of an agent, in this case only given by the} Integrated Information, with its Morphological Computation. Here the concept of \emph{Morphological Computation}  describes the reduction of computational cost for the controller that results from the interaction of the agent's body with its environment. One example where Morphological Computation is applied is the field of soft robotics. There the softness of the robots' bodies leads to a lower computational cost when they, for example, grab fragile objects \citep{Ghazi-Zahedi2017b, Kohei, Kohei2014}. Different understandings of Morphological Computation are discussed by \citet{Mueller, Ghazi-Zahedi2019}.  \citet{Auerbach} analyze simulated evolving agents and conclude that the complexity of the morphology of an agent depends on its environment. In the field of embodied artificial intelligence the cheap design principle, formulated by \citet{Pfeifer2006}, states that a robot's body should be constructed in a way that best exploits the properties of the environment. This should lead to a simpler control architecture. \textcolor{changed}{The cheap design principle is discussed in the context of universal approximations in \citep{Montufar2015}.}

We confirm this intuition in \citep{Langer2021} \textcolor{changed}{in experiments with simulated agents}, where the comparison between the controller complexity and Morphological Computation leads to the result that they are inversely correlated. On the one hand this is intuitive, since the more the agent relies on the interaction of its body with the environment to solve a task, the less involvement of the controller is needed. On the other hand this leads to the problem that now embodied intelligence is correlated with reduced involvement of the brain. If the morphology of an agent's body is intelligent enough, would it need a control architecture at all?

Here we want to present an additional perspective by considering the challenge of learning. As discussed in the introduction, learning to perform a task entails updating an internal world model in order to predict the outcome of ones actions. Hence, we measure the controller complexity not only via the Integrated Information but also by the complexity of the internal world model. We hypothesize that a learning process requires the agent to highly integrate the available information, hence that learning requires an increased controller complexity. \citet{Edlund} conclude that Integrated Information increases with the fitness of evolving agents. \citet{Albantakis2014} increase the complexity of the environment, which leads to higher Integrated Information and \citet{Albantakis2015} observe that high Integrated Information benefits rich dynamical behavior. All these results are clear indications that a high information integration in the controller is beneficial for an embodied agent that is faced with a task.

\textcolor{changed}{Note that the complexity measures in the context of Integrated Information focus on the mechanistic structure of the information flow inside the controller and the internal world model, not the actual quality of the internal world model. An alternative perspective on assessing an internal model is to measure how much of the environmental states the internal model captures, how much of the environment it represents, giving rise to internal representations. In \citep{Ashby} the author postulates that the number of internal states in a controller needs to be greater or equal to the system being controlled in order to be stable, hence this defines a lower bound on the size of a representation. The importance of predicting the next sensory states via an internal model is discussed in, for instance, \cite{Clark2016}.}

\textcolor{changed}{
The necessity of representing the environment for an artificial agent was called into question by \cite{Brooks1991}. This point of view and further criticism towards the representationalist approach are discussed in detail in \citep{Clark1994}. A thorough introduction to the history of representations can be found in \citep{Marstaller2013},}
\textcolor{changed}{ where the authors define the representation explicitly as the information about the environment encoded in the internal states which goes beyond the information in the sensors. They show that this measure increases with the fitness of simulated agents equipped with an evolutionary algorithm. Aside from the different type of measure our approach focuses on understanding how individual agents learn a new task in their lifespan, not on a population level.    }

Using the simulated, \textcolor{changed}{learning} agents we first consider the results of a type that do not need to form an internal world model, the \PWM agents. \textcolor{changed}{These agents have access to a sampled, external world model, which describes their experiences instantaneously and accurately.} In this utopian situation agents do not require a complex controller in order to learn and they behave mainly through reactive mechanisms, as long as their world model is accurate. In contrast, the  \LWM  agents, the ones that have to \textcolor{changed}{learn} their internal world models, require an increased controller complexity in order to \textcolor{changed}{successfully} learn. Once their world model is accurate, the Integrated Information value decreases, since the agent can then make use of the interactions with its environment, measured by Morphological Computation.

\textcolor{changed}{We summarize the intuition and main results of our experiments in the next section.}
%Additionally, we take a close look at the dynamics of the prediction process \textcolor{changed}{using a synergistic measure on the internal world model, which is part of the controller complexity}. We compare the complete agents with ones that have a simplified controller. These agents are not able to integrate information in their controller and are therefore called \emph{split}. We see that the latter are barely, if at all, able to learn to perform the desired task. The ones that are successful have a complex prediction mechanism in their internal world model.

\subsection*{Intuition and Main Results}

Learning a new task and adapting to changes in the world poses a difficult challenge. An important aspect of this is to predict the outcomes of the own actions.  We theorize that even for seemingly easy situations, in which agents can manage without much involvement of the brain, learning the best behavior requires complex computations in the controller. We illustrate this in the following example.

Consider a child learning to ride a bike. Nearly every task the child has learned previously, e.g.~walking, speaking or drawing, becomes harder when one tries to do it fast. So the child expects that moving slowly leads to the best outcome. According to its understanding of the world, its world model, riding a bike slowly is easier than doing it fast. Unfortunately speed is required to stabilize a bike. The child is working with an inaccurate world model. So before the child can learn to ride a bike it has to \textcolor{changed}{assess the information from its experiences and} understand that faster can mean easier. It has to update its world model to learn and to be able to use the world in an optimal way.

\begin{figure}[h]
    \centering
    \includegraphics[width=0.55\textwidth]{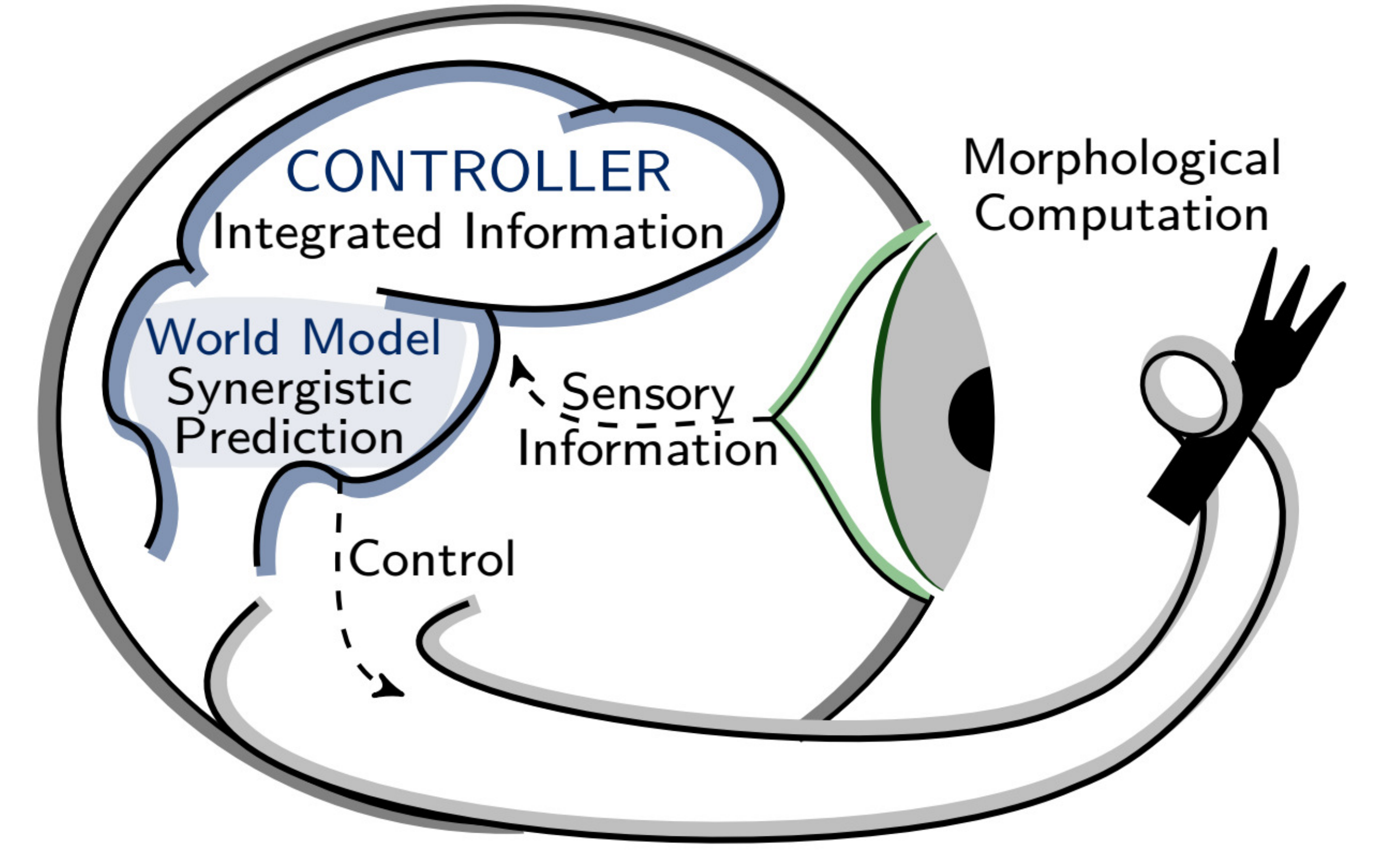} 
    \caption{Sketch of an agent interacting with its environment and the different measured information flows.}
    \label{fig:alien}
\end{figure}

\textcolor{changed}{In order to analyze these dynamics, } we closely examine the information flows in learning agents.  \textcolor{changed}{In Figure \ref{fig:alien} we depict a sketch of an agent interacting with its environment, a fork, and highlight the different information flows that we analyze in this work. The agent perceives its environment through its eye and we quantify the importance of the information flow from the sensors to the controller by a measure called \emph{sensory information}. We assess the complexity of the controller by two different measures, namely \emph{Integrated Information} and \emph{synergistic prediction}, both are described in more detail below. The information flow from the controller to the actuators, which then determine the actions of the agent, is measured by \emph{control}. Lastly, the interaction among an agents body and the environment, which reduces the computational cost for the controller, is called Morphological Computation. In the sketch this is given by an octopus-like arm holding a fork.}

We are especially interested in the controller complexity, quantified by two measures that assess distinct mechanisms of the controller.
The first measure can be seen in the context of the Integrated Information Theory of consciousness, introduced in the previous section, and it quantifies the information integration among parts of the controller. \textcolor{changed}{Additionally,} the second measure assesses the complexity of the agent's internal world model, which predicts the next sensory state. 
\textcolor{changed}{It is called \emph{synergistic prediction}. Both measures follow the notion that a system is complex, if it is more than the sum of its parts.}

%We theorize that an agent has to combine its available information, either by using Integrated Information or a complex internal world model, in order to learn.

We use simple simulated agents and observe how the complexity of the controller develops during the learning process. Our first conclusion can be summarized as follows:

\begin{enumerate}
%\item[1.] The information from the environment needs to get %integrated in at least one point in the agent. 
\item\label{itm:first} An agent that understands its environment, meaning it has an accurate world model, exhibits a higher Morphological Computation and lower controller complexity compared to agents with an inaccurate world model. The better an agent understands its environment, the more it can exploit the interactions between body and environment and the less controller complexity is needed.  
\end{enumerate}

This \textcolor{changed}{conclusion} is supported by the following observations. At first we analyze agents that do not have to learn \textcolor{changed}{an internal} world model.  \textcolor{changed}{Instead, each of these agents is able to access an external world model, which samples the dynamics of the immediate environment of the agent. This then accurately describes the agents experiences, it is ideal}, hence we refer to these agents as \PWM agents. We observe that they need next to no involvement of the controller and that the interaction with the environment, referred to as Morphological Computation, increases with the accuracy of the world model. At the same time the influence of the controller on the behavior of the agents is high for an inaccurate world model and decreases as the quality of the world model improves. 

Furthermore, we refer to agents that have to learn an internal world model as \LWM agents. They initially have a high controller complexity and then this value decreases, if they are successful. Hence, we conclude above that the agents first have to learn the correct world model, before they are able to optimally utilize the interaction of their bodies with the environment, which in turn leads to a lower controller complexity. Moreover, this theory is supported by the result that unsuccessful agents have a constantly high controller complexity and a lower Morphological Computation compared to the successful agents.

Additionally, we analyze agents with a simplified control architecture for which the ability to integrate information is \textcolor{changed}{inhibited}. Hence, \textcolor{changed}{the controller of these agents is divided into two unconnected parts,} they have an Integrated Information of zero and we call them \emph{split} \LWM agents. The controller complexity of these agents is solely determined by the second measure, assessing the internal world model, and they perform noticeably worse compared to  complete \LWM agents. The few successful split \LWM agents have a complex internal world model, which leads us to the following conclusion:

\begin{enumerate}
\setcounter{enumi}{1}
\item\label{itm:second} In order to successfully learn the agents have to combine information from different sources. This leads to an increased controller complexity either in form of Integrated Information or in the prediction process given by the internal world model.
\end{enumerate} 

In the next section we introduce the experiments and the agents in more detail. % \ref{itm:second}

\section*{Materials and Methods}
\subsection*{Setting of the Experiment} \label{Sect:Setting}
\begin{figure}[htp] % Example Figure
    \centering
    \begin{minipage}[c]{0.425\textwidth}
    \includegraphics[width= \textwidth]{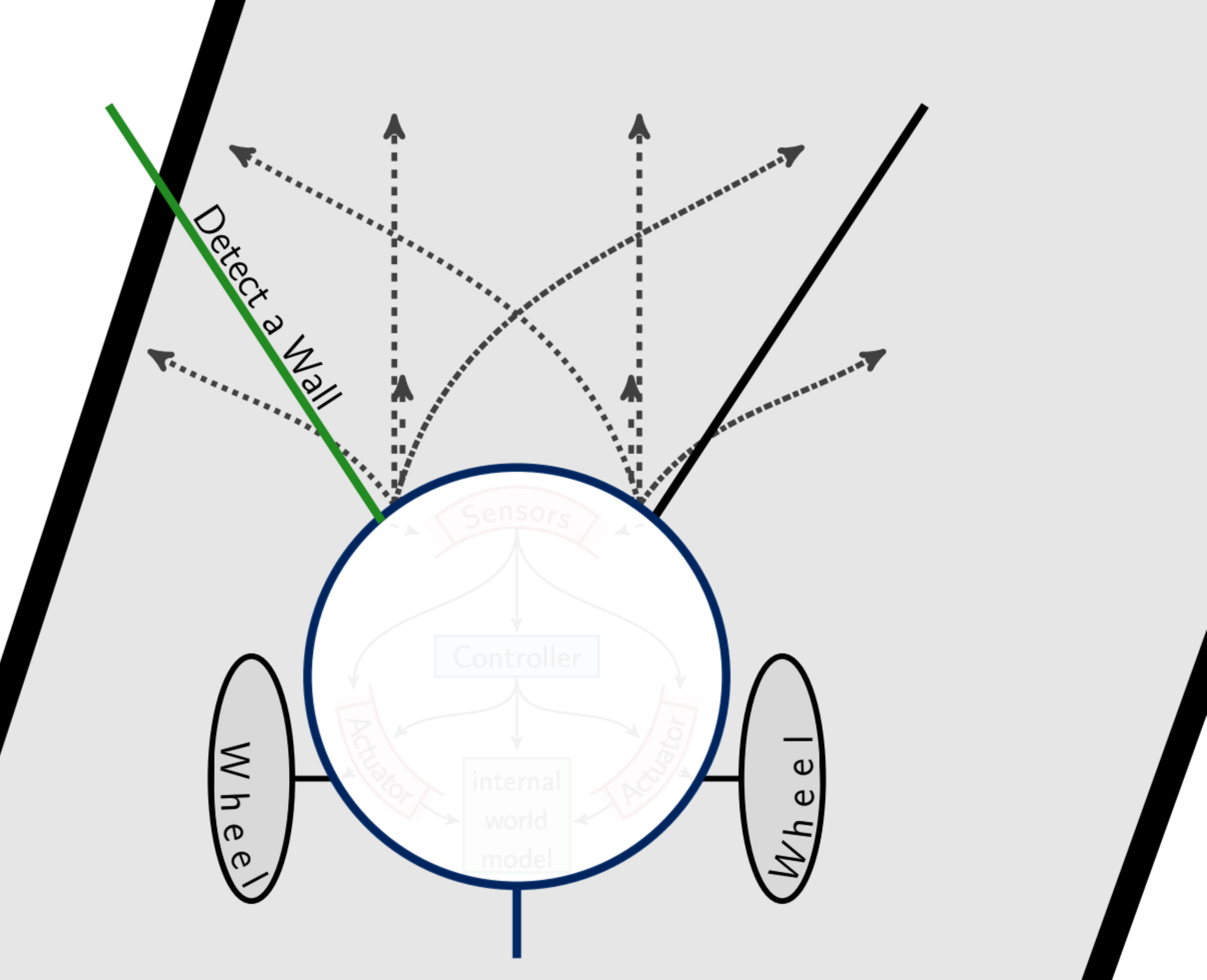}  
    \end{minipage} \hfill 
    \begin{minipage}[c]{0.425\textwidth}
    \includegraphics[width= \textwidth]{Figure2b-eps-converted-to.pdf}
    \end{minipage}
    \caption{\textcolor{changed}{Sketch of a two-wheeled agent and its four movements in its environment on the left. }Five different agents in their environment in the middle and the possible sensor length from 0.5 on the top right to 2 on the bottom right.}
    \label{fig:1}
\end{figure}

In our experiment we analyze the information flows of simplistic, 2-dimensional, acting agents. An agent consists of a round body \textcolor{changed}{with a radius of 0.55 unit length}, a small tail and two binary sensors. \textcolor{changed}{The tail simply marks the back of the agent and has no influence on its behavior.} Two range sensors are visualized in Figure \ref{fig:1} on the left as lines that are green when they detect a wall and black otherwise.  \textcolor{changed}{We vary the reach of these sensors, as discussed in more detail below. The agents can be thought of as two-wheeled robots, as sketched on the left of Figure \ref{fig:1}. Each wheel can spin either fast or slow, which leads to} four different movements, which are fast forward (approx. $0.6$ unit length per step), slow forward (approx. $0.2$ unit length per step), left and right (with approx. $14^{\circ}$ and a speed of $0.4$ unit length per step).

Five of these agents are depicted \textcolor{changed}{in the middle of Figure \ref{fig:1}} in the racetrack in which they have to move.
Whenever the body of an agent touches a wall, the agent gets stuck. This means that it can only turn on the spot and will not be able to move away unless both sensors do not detect a wall anymore. The implementation and a video of this movement can be found at \citep{Langer2022}.

\textcolor{changed}{Additionally, we vary the length of the sensors from 0.5, depicted on the top right of Figure \ref{fig:1} to a sensor length of 2, shown in the bottom right of the figure, with increments of 0.25.}
Varying the length of the sensors directly influences the amount of information an agent receives about the world and hence it can influence the quality of the interaction of the agent with its environment. Therefore this has an impact on the potential Morphological Computation. \citet{Mueller} call this \emph{Morphology facilitating perception} and discuss  \textcolor{changed}{its  relationship to other types of Morphological Computation in more detail}.

\subsection*{The Agents and the World Models} \label{Sect:Agents}

\begin{figure}[htp]
    \centering
    \vspace{1cm}
    \begin{minipage}[c]{0.3\textwidth}
    \includegraphics[width=1\textwidth]{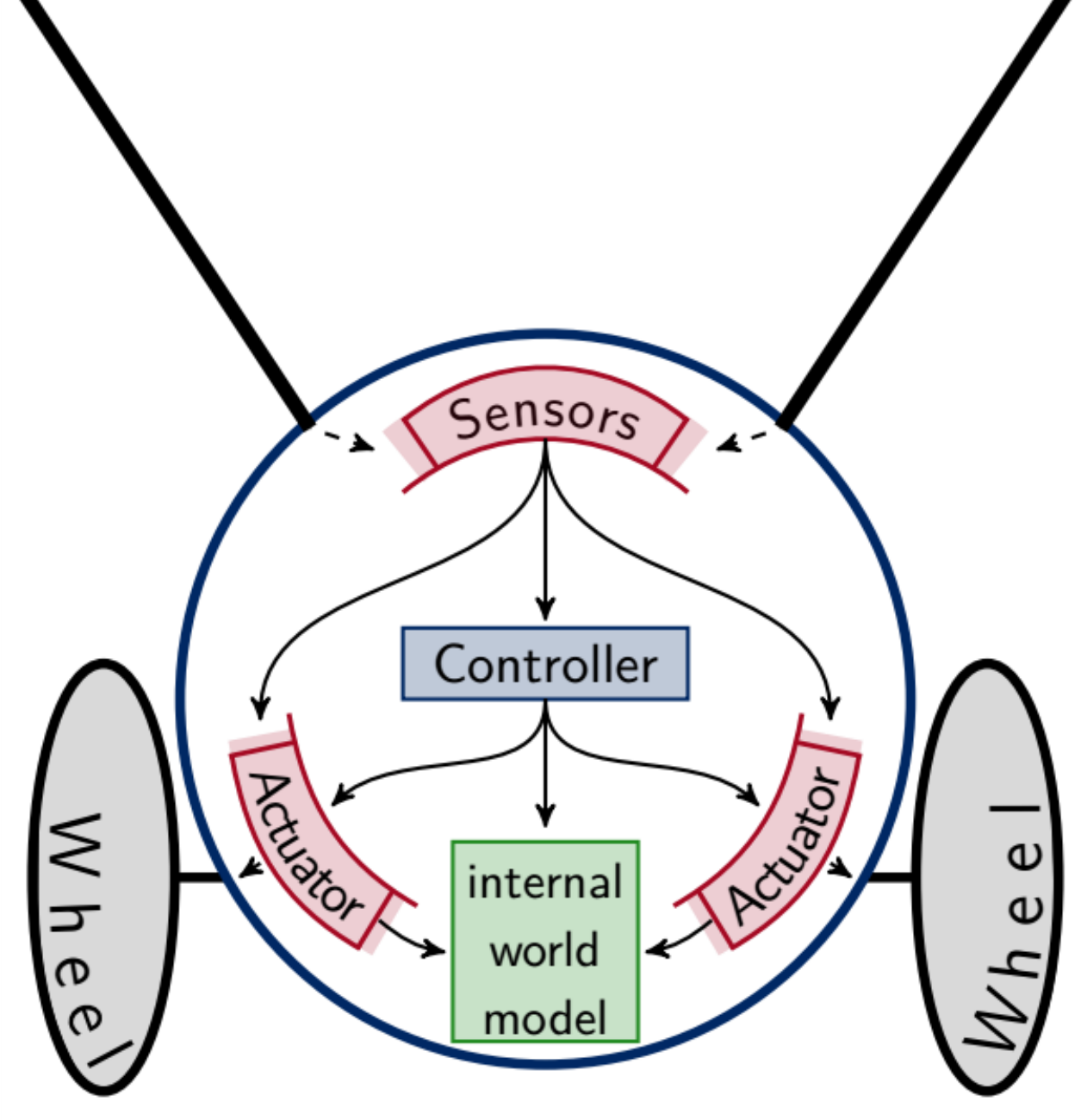}  \hfill
    \end{minipage} \hfill
        \begin{minipage}[c]{0.65\textwidth}
    \includegraphics[width=1\textwidth]{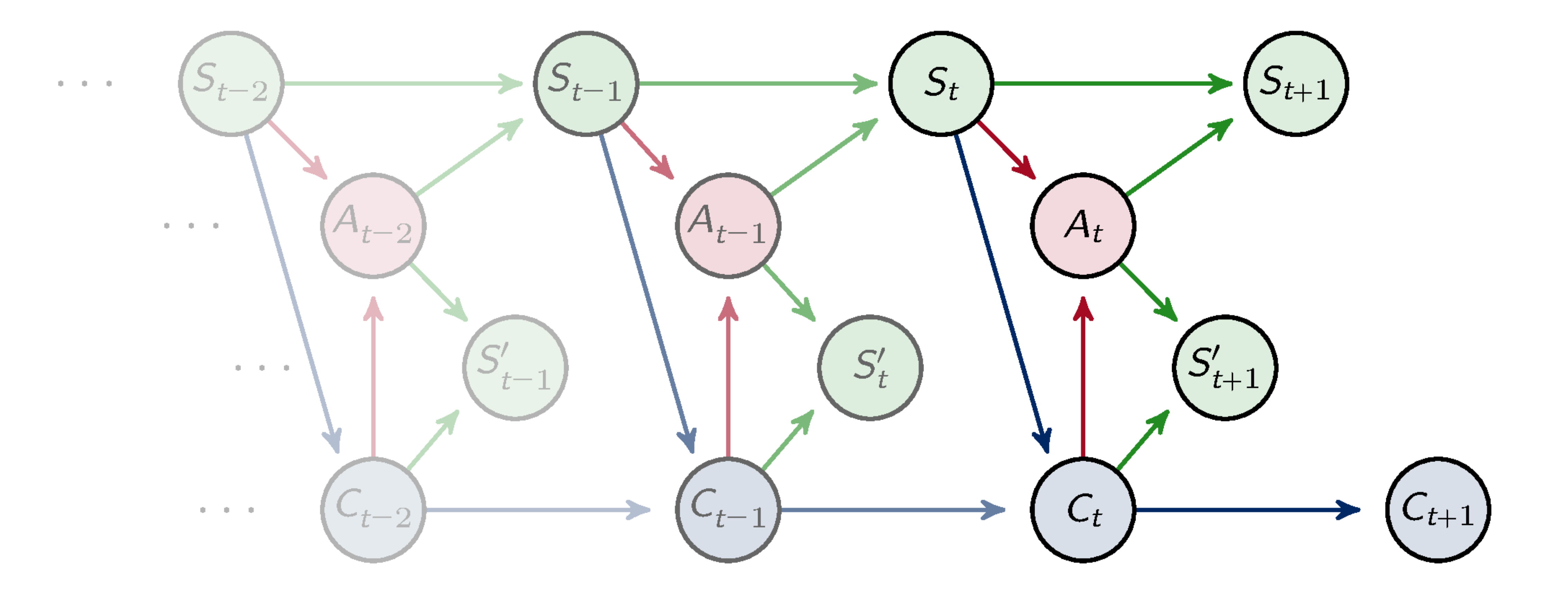}
    \end{minipage}
    \caption{Sketch of the architecture of an \LWM agent on the left and the sensorimotor loop of this agent on the right.}
    \label{fig:2}
\end{figure}

An agent is modeled by a discrete multivariate, time-homogeneous Markov process, denoted by
$(X_t)_{t\in \mathbb{N}} = (S_t, A_t, C_t)$
with the state space $\mathcal{X} = \mathcal{S} \times \mathcal{A} \times \mathcal{C}$. 
The variable $S_t$ entails the two binary sensors that detect a wall and a binary variable encoding whether the agent is touching a wall. The node $A_t$ includes two binary actuators and $C_t$ two binary controller nodes. Additionally, \textcolor{changed}{in case of the \LWM agents} the variable $S'_t$ describes the internal prediction of the next sensor state and hence consists of three binary variables. The connections among these variables are sketched on the left of Figure \ref{fig:2}. %The \PWM agents do not have an $S'_t$. 

The elements of the \LWM agents are connected according to the graph in Figure \ref{fig:2} \textcolor{changed}{on the right}. An introduction to graphical models is given by \citet{Lauritzen}. We depict only one node for each $S, S', A$ and $C$ in the figures to increase clarity.  The factorization of the corresponding probability distribution is given in  \eqref{eq:fact}
\begin{equation}
\begin{split}
P(X_t, X_{t\scalebox{0.65}{+}1}, S'_{t \scalebox{0.65}{+}1}) = P(S_t, A_t, C_t) &P(S'_{t\scalebox{0.65}{+}1} \vert A_t, C_t) P(S_{t\scalebox{0.65}{+}1} \vert S_t, A_t) \\ &\prod\limits_{j=1}^2 P(C^j_{t\scalebox{0.65}{+}1} \vert C_t, S_{t\scalebox{0.65}{+}1}) \prod\limits_{i=1}^2 P(A^i_{t\scalebox{0.65}{+}1} \vert S_{t\scalebox{0.65}{+}1} , C_{t\scalebox{0.65}{+}1} ). \label{eq:fact}
\end{split}
\end{equation}
Since $S_t'$ is a prediction of $S_t$ it is made of the same substrate, hence the state space of $S_t'$ is $\mathcal{S}$. The difference between $S_t$ and  $S_t'$ lies in the mechanism with which they are generated. The node $S_t$ is influenced by the information from $S_{t-1}$ and $A_{t-1}$. These are indirect influences, because in this case the information flows through the environment. The role of the environment is discussed in more detail in \citep{Langer2021}.

The conditional distribution, $P(S_{t+1} \vert S_t, A_t)$, is called \textit{world model} by  \citet{Ghazi-Zahedi2010, Montufar2015}.
The internal prediction $S_t'$ is generated by $P(S'_{t+1} \vert A_t, C_t)$ and also named \emph{world model} by \citet{Ay2013, Ay2013a, Ay2014}. To prevent confusion we refer to $P(S'_{t+1} \vert A_t, C_t)$ as \emph{internal world model} and to $P(S_{t+1} \vert S_t, A_t)$ as \emph{ideal world model}. We use the latter term, since this distribution is defined by sampling the \textcolor{changed}{ individual past experiences of the agents, hence the ideal world model always represents the experiences of the agents accurately}.

\begin{figure}
    \centering
    \includegraphics[width=\textwidth]{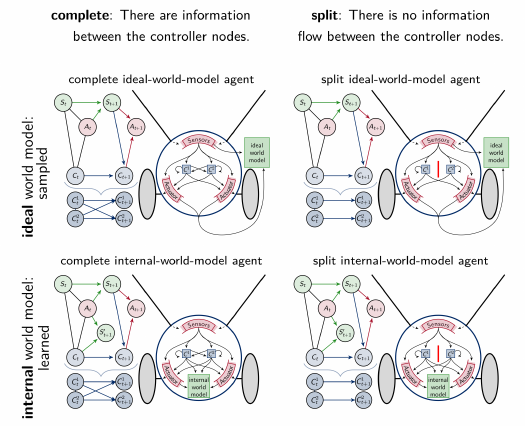} 
    \caption{The connections of the complete
and split \PWM agents on the top and
the complete and split \LWM agents on the bottom.}
    \label{fig:3}
\end{figure}

\textcolor{changed}{In total we analyze the behavior of four types of agents, summarized in Figure \ref{fig:3}. The first distinction among the agents is between those with a complete controller, depicted on the left of Figure \ref{fig:3}, and agents with a simplified controller, which are called split agents and are depicted on the right of Figure \ref{fig:3}.
The latter agents are not able to integrate information between the controller nodes, since the controller node $C_{t+1}^i$ only receives information from $C_{t}^i$ and not from $C_{t}^j$, $i,j \in \{1,2\}, i \neq j$.}

\textcolor{changed}{
 Secondly, we differentiate between agents with and without an internal world model. The agents on the top of Figure \ref{fig:3} have no internal world model. These agents have direct access to their sampled, ideal world model and they are called \PWM agents, whereas the \LWM agents, depicted on the bottom of Figure \ref{fig:3}, have to learn their internal world models.  }

\subsection*{Learning}

In our previous publication \citep{Langer2021} we use the concept of \emph{Planning as Inference} in order to optimize the behavior of the agents and we will apply the same algorithm here in the case of the \PWM agents. In this method the conditional distributions are optimized with respect to a goal variable by using the em-algorithm. This is a well-known information geometric algorithm that is guaranteed to converge, but might converge to a local minimum \citep{Amari1995, Amari2007}. 
\textcolor{changed}{This algorithm minimizes the difference between two sets of probability distributions by iteratively projecting onto them.}

\textcolor{changed}{
In the case of the \LWM  agents we have two goals. Firstly, we want to optimize the distributions determining the behavior, $P(C_{t+1} \vert  S'_{t+1}, C_t)$ and $ P(A_{t+1} \vert S'_{t+1} , C_{t+1} )$, such that the probability of touching the wall after the next movement is as low as possible. At the same time, the internal world model, $P(S'_{t+1} \vert C_t, A_t)$, should be close to the actual, ideal world model, $P(S_{t+1} \vert S_t, A_t)$.} This second goal is important, because otherwise the optimization of the behavior would use faulty assumptions leading to a failure of the agent. In the example in the introduction this would be the child trying to learn to ride a bike, while going as slowly as possible. Hence, both of the world models should result in similar predictions.

So we modify the em-algorithm to alternate between optimizing \textcolor{changed}{the agent} with respect to the goal on one hand and with respect to the difference between \textcolor{changed}{world models} on the other hand. Details of this optimization are described in the Appendix \ref{S1_Appendix}. 

Note, that the controller has only two binary variables, whereas $S$ consists of three. Therefore merely copying the information from the sensors is not a viable strategy for the agents. \textcolor{changed}{Even though we are studying simple agents here, this is a natural setting, compared with human perception. We, as humans, do not consciously perceive every detail of our environment, instead we learn to distinguish between important and irrelevant information.} % because we are not able to consciously perceive every detail of our environment that our sensors pick up, instead we learn to distinguish between important and irrelevant information. 

\subsection*{Measures of the Information Flow} \label{Sect:Meas}
We measure the importance of an information flow by calculating the difference between the actual distribution and the closest distribution without the information flow in question. In Figure \ref{fig:6} we emphasize the measured connection by a dashed arrow. The set of distributions without this information flow is called a split system. 
More precisely, the measure in case of a split system $M$ is defined in the following way. 
\paragraph{Definition} \label{Def:Psi}
Let $M$ be a set of positive probability distributions on a state space $\mathcal{Z}$, referred to as split system.
Then we define the measure $\Psi_{M}$, by minimizing the Kullback-Leibler divergence between the split system $M$ and the full distribution $P$
\begin{equation}
\Psi_M = \inf\limits_{Q \in M} D_{\mathcal{Z}}(P \parallel Q) = \inf\limits_{Q \in M} \sum\limits_{z \in \mathcal{Z}} P(z) \, log \, \dfrac{P(z)}{Q(z)}. \label{psi}
\end{equation} 
Most of the discussed measures have a closed form solution and can be written in the form of sums of conditional mutual information terms. The conditional mutual information,  $I(Z_{1} ; Z_{2} \vert Z_{3})$, is defined in \eqref{eq:mutualinf} 
\begin{equation}
\begin{split}
 I(Z_{1} ; Z_{2} \vert Z_{3}) = \sum\limits_{z_1, z_2, z_3} P(z_1 , z_2,z_3) \, log \, \dfrac{P(z_1 , z_2 \vert z_3)}{P(z_1 \vert z_3) P(z_2 \vert z_3)} \label{eq:mutualinf}
\end{split}
\end{equation}
and can be interpreted as follows.
If $I(Z_{1} ; Z_{2} \vert Z_{3}) = 0$, then $Z_{1}$ is independent of $Z_{2}$ given $Z_{3}$. Hence, this quantifies the connection between $Z_{1}$ and $Z_{2}$, given the influence of $Z_{3}$. 

\newpage
\subsubsection*{Controller Complexity}
We assess the controller complexity using two different concepts that refer to different parts of the controller. First we discuss the measure corresponding to the Integrated Information, before we quantify the complexity of the internal world model.
\paragraph{Integrated Information}
There exist various types of Integrated Information measures, as discussed in the introduction.
The approach we are using here was defined in our previous publication \citep{Langer}, also applied in \citep{Langer2021}. 
There we quantify how much information gets integrated among different controller nodes across different points in time, as depicted in Figure \ref{fig:6} (A).
The minimization described in \eqref{psi} results in the solution below, as shown in \citep{Langer}. 
\vspace{-0.25cm}
\begin{equation}
\Phi_{IIT} = \sum\limits_j I( C_{t+1}^j ; C_t^{J \setminus \{j\}} \vert C_t^j , S_{t+1})
\end{equation}
In our case we only have two binary controller nodes, hence $J = \{1,2\}$. 
Note that the split  agents do not have these connections leading to $\Phi_{IIT} = 0$.  
%\LWM and \PWM

The importance of the Integrated Information for the behavior of an agent also depends on the information flowing to and from the controller, as observed in \citep{Langer2021}. This is quantified by the two following measures, namely sensory information and control.

We assess the importance of the information flow from the sensory nodes to the controller nodes by a measure called \emph{sensory information}. The graphical representation of the split system is depicted in Figure \ref{fig:6} (B) and the closed form solution of this measure is
\vspace{-0.25cm}
\begin{equation}
\Psi_{SI} = \sum\limits_j I(C^j_{t+1} ; S_{t+1} \vert C_t ).
\end{equation}
If this value is zero, then the controller nodes do not depend on the sensory input and therefore cannot make any behaviorally beneficial contributions.

Additionally, the strength of the connection from the controller nodes to the actuator nodes is assessed by a measure that we call \emph{control}, $\Psi_{C}$
\begin{equation}
\Psi_C = \sum\limits_i I(A_{t+1}^i ; C_{t+1} \vert S_{t+1}).
\end{equation}
An agent in which its controller has no influence on the actuator at all has $\Psi_C = 0$. 

The combination of these three measures is an indicator of the impact that the Integrated Information has on the behavior of the agent. This is called \emph{effective information integration} and defined as the product: 
\begin{equation}
\Phi_{EII} = \Phi_{IIT} \cdot \Psi_{SI} \cdot \Psi_{C}. \label{phi_eii}
\end{equation} 
\paragraph{Internal world model} \label{Subsubsect:Pred}
We analyze the internal world model $P(S_{t+1}' \vert C_t, A_t)$ by calculating how important the interplay between the influences of $A_t$ and $C_t$ on $S'_{t+1}$ is.

This measure has no closed form solution. Here we define a split system $Q$ as consisting of only the two-way interactions among the three variables, namely $Q(A_t,C_t), Q(A_t,S'_{t+1})$ and $Q(C_t, S'_{t+1})$, but without combined influence from $(A_t, C_t)$ on $S'_{t+1}$.
Hence, we call this measure synergistic prediction,
$\Psi_{SynP}$. The two-way interactions are highlighted in Figure \ref{fig:6} (C).
This is conceptually similar to the synergistic measure for Morphological Computation proposed in \citep{Ghazi-Zahedi2017a} and we also use the iterative scaling algorithm to calculate this measure, as described there in Section 2.5. %in \citep{Ghazi-Zahedi2017a}. 

%\begin{figure}[hpt]
%\centering
%\includegraphics[width = 0.2\textwidth]{Figures/Figure8.eps}
%\caption{Sketch of the split system in case of $\Psi_{SynP}$.}
%\label{fig:8}
%\end{figure}

\subsubsection*{Morphological Computation}
\begin{figure}[ht!]
\centering
\includegraphics[width = \textwidth]{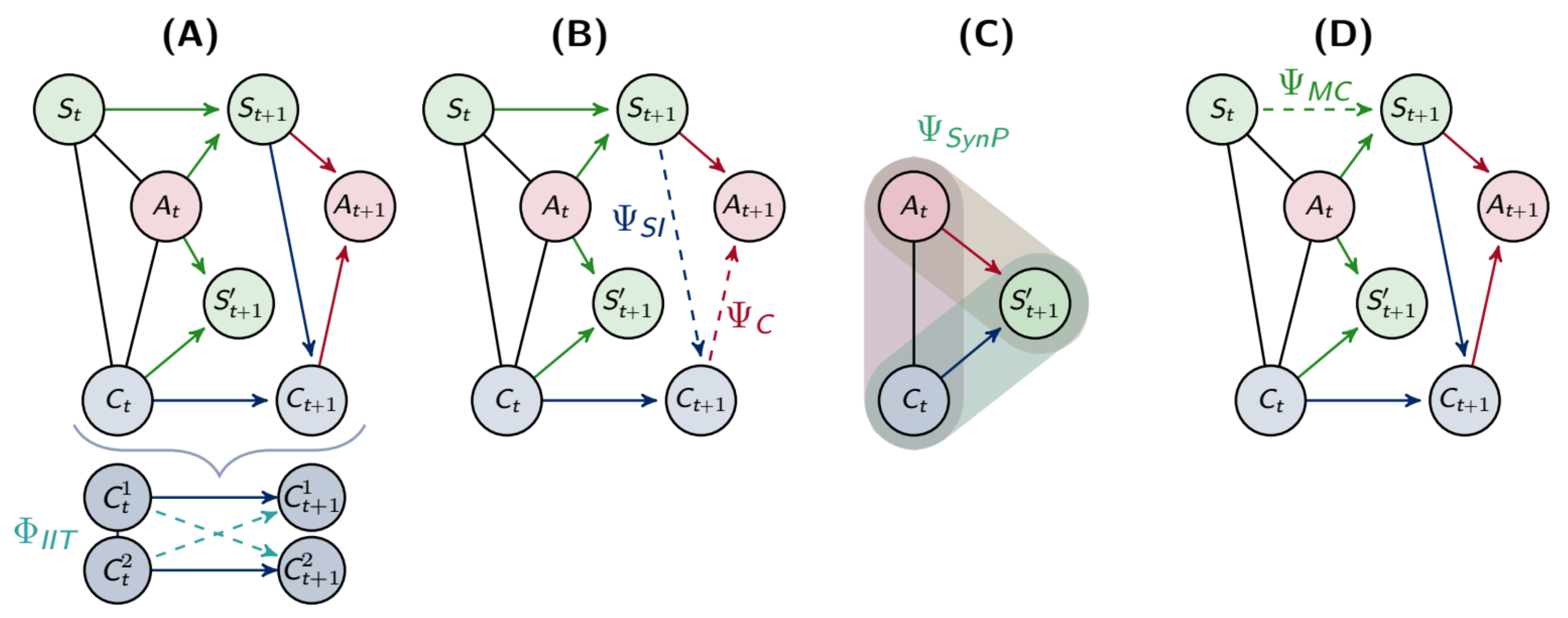} 
\caption{Graphs corresponding to the split systems in case of $\Phi_{IIT}$ in (A), $\Psi_{SI}$ and $\Psi_C$ in (B), $\Psi_{SynP}$ in (C) and $\Psi_{MC}$ in (D).}
\label{fig:6}
\end{figure}

The concept of Morphological Computation describes the reduction of the necessary computation in the controller that results from the interaction of the agent's body with its environment. There exist various types of Morphological Computation and different measures for it \citep{Mueller}. We  use the following formulation
\vspace{-0.25cm}
\begin{equation}
\Psi_{MC} = I(S_{t+1} ; S_t \vert A_t).
\end{equation}
This measures the information flow  going from one sensory state to the next one through the world, given the actuator state, as depicted in Figure \ref{fig:6} (D). In \citep{Ghazi-Zahedi2013} this was introduced as a measure for Morphological Computation and in \citep{Ghazi-Zahedi2019} in a comparison with other measures $\Psi_{MC}$ shows desirable properties.

%\begin{figure}[hpt]
%\centering
%\includegraphics[width = 0.3\textwidth]{Figures/Figure9.eps}
%\caption{{Graph corresponding to the split system in case of $\Psi_{MC}$.}}
%\label{fig:9}
%\end{figure}

%
%
%\begin{minipage}[t]{0.75\textwidth}
%\setlength{\parindent}{1.5em}
%Reactive Control
%\begin{align*}
%\Psi_R = \sum\limits_i I(A^i_{t+1} ; S_{t+1} \vert C_{t+1})
%\end{align*}
%\end{minipage} \hfill
%\begin{minipage}[t]{0.25\textwidth}
%  \centering\raisebox{\dimexpr \topskip-\height}{%
%\includegraphics[width = \textwidth]{LearningReactiveControl} }
%\captionof{figure}{Graph corresponding to the split system ... }
%\label{SplitReactive} 
%\end{minipage}
\section*{Results}
In this section we discuss the results of our simulations. We used 1000 random input distributions for each sensor length and each type of agent. All agents train for 20000 steps and the measures are calculated for 90 different points during these steps. More precisely, we apply the measures for the 9 time points, namely 
50, 100, 200, 500, 1000, 2000, 5000, 10000, 20000, 
and additionally 9 equidistant time points between each of them, as well as 9 equidistant time points between zero and 50.

Additionally, we calculate the success rate (SR) of an agent by sampling how many time points the agent is stuck at a wall during the 20$\,$000 training steps. \textcolor{changed}{Hence an SR of $0.1$ signifies that an agent was not stuck $10 \%$ of the steps.}
We then divide the agents in successful and unsuccessful ones based on their SRs.  
The best third of the complete \LWM agents perform above $16.8\%$ and are called successful, while we refer to agents with a SR below $16.8\%$ as unsuccessful. Dividing the agents in this way allows us to call only the agents successful for which the success rate increased noticeably during learning. 

\begin{figure}[htp]
\centering
\includegraphics[width = 0.925\textwidth]{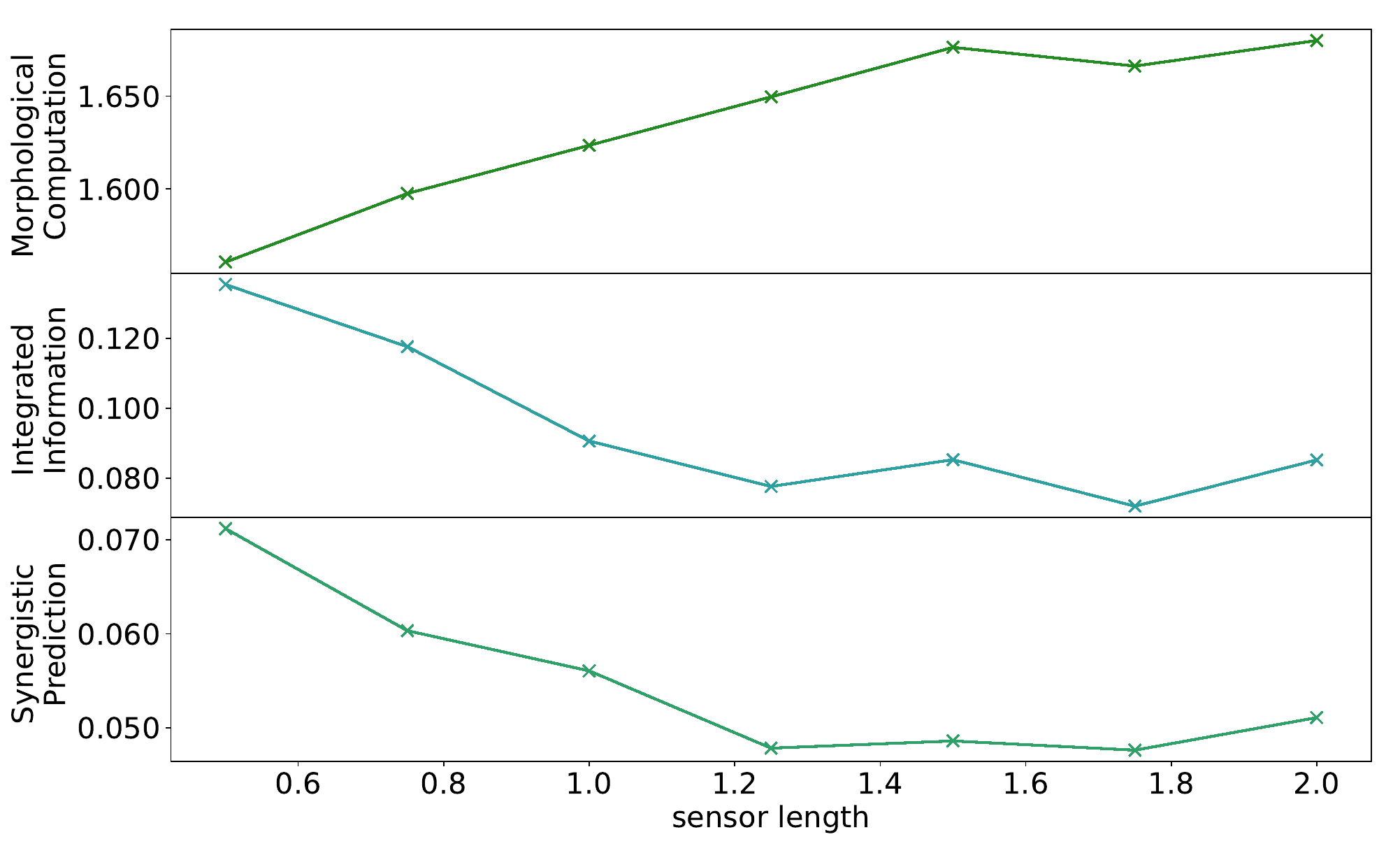}
\caption{{Morphological Computation, Integrated Information and Synergistic Prediction for the successful,
complete \LWM agents.}}
\label{fig:10}
\end{figure}

In Figure \ref{fig:10} we can see the results of the measures for the controller complexity, namely Integrated Information and synergistic prediction, as well as the Morphological Computation averaged over all successful  \LWM  agents after 20$\,$000 steps. 

We observe that the controller complexity and the Morphological Computation are inversely correlated. Therefore the results confirm our previous observation, published in \citep{Langer2021}, \textcolor{changed}{that Morphological Computation and Integrated Information have an inverse relationship}. 
Note, that when the sensors are too long so that the agents almost always detect a wall this additional information is no longer beneficial for the agents and the Morphological Computation no longer increases, while the Integrated Information and synergistic prediction increase again.

This relationship leads to the question, why agents with a well-adapted morphology would need a complex control architecture. Wouldn't it be possible to build agents that are so well adapted to their environment that a simple controller suffices? There might be several reasons why a complex controller  is necessary in general, despite this inverse correlation, as we discuss further in the section \enquote{Conclusion}.  

Here we argue that an involvement of the controller is necessary, since agents first have to learn how to interact with their environment, meaning they have to build their world models.
\subsection*{The Ideal-World-Model Agents} \label{Subsect:Ideal}

\begin{figure}[htp]
\includegraphics[width = \textwidth]{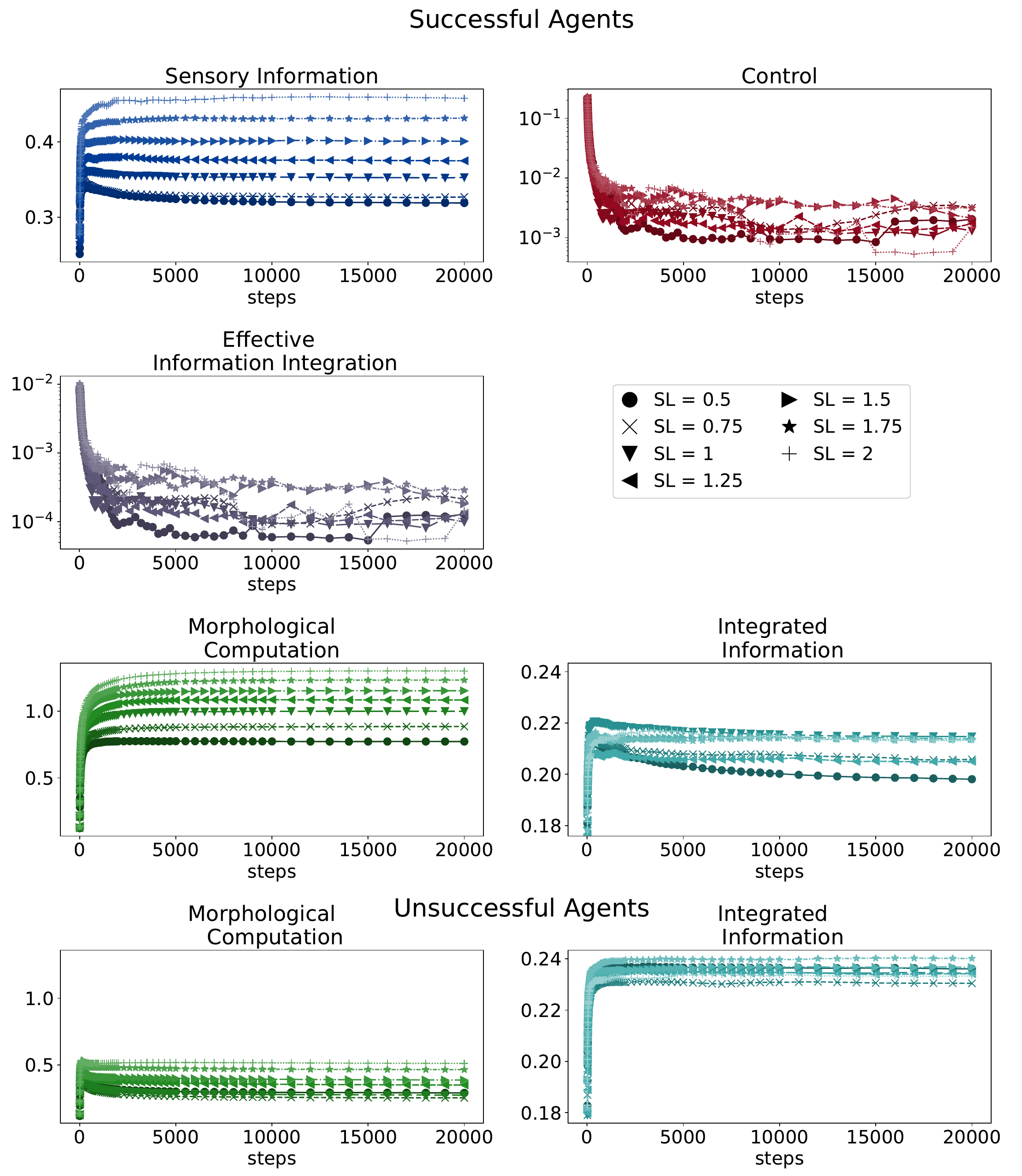} 
%\vspace{-1cm}
\caption{{The sensory information, control, effective information integration, Morphological Computation and Integrated Information for the
successful, \PWM agents in the first three rows and the Morphological Computation and Integrated Information for the unsuccessful \PWM agents in the last row.}}
\label{fig:11}
\end{figure}

Now, we discuss the results for the \PWM agents that do not have to learn their world models, \textcolor{changed}{since they have direct access to the sampled ideal world models.} The best approx.
$33\%$  of the \PWM agents are the ones with a success rate above $61,5\%$, which we term the successful \PWM agents. Hence, the \PWM agents perform overall much better than the \LWM 
agents, for which the best third only performs better than $16.8\%$. We depict the Integrated Information, sensory information, control, effective information integration and \textcolor{changed}{Morphological Computation for the successful \PWM agents in the first three rows of Figure \ref{fig:11}.} 

%\vspace{-0.5cm}

The controller complexity, given here solely by the Integrated Information value due to the lack of an internal world model, seems to not change after the first few initial steps. In \citep{Langer2021} we discuss that the importance of the controller complexity additionally depends on the sensory information and the control. While the sensory information increases with the sensor length, we can see the reason for the behavior of the Integrated Information in the results of $\Psi_C$, \textcolor{changed}{the measure for control}. After the first steps this measure is very close to 0 with an average value of $0.0021$ at the 20$\,$000th step. If $\Psi_C = 0$, then the controller has no influence on the behavior of the agent at all.  It is easy to check that in this case the information flow in the controller is not changed by the em-algorithm anymore, since the controller has no influence on whether the agent is successful or not. This only holds for the \PWM agents, because we apply the original em-algorithm here, not the modified one.

The effective information integration, \textcolor{changed}{ depicted in the second row and first column} of Figure \ref{fig:11}, summarizes the behavior of the other three measures. This has a value close to zero which shows that the controller complexity is nearly irrelevant
for the behavior of the agent in this case.

Hence, for the \PWM agents a complex controller is not needed in order to learn to perform a task. In fact, split \PWM agents, without the ability to integrate information, perform only slightly worse than complete ones. More precisely, the split \PWM agents have an average success rate of $33.69\%$ compared to $33.83\%$ in the complete case. 

In this scenario, the success does not depend on the complexity of the controller, but on the interaction of the agent with its environment. We therefore now directly compare the Morphological Computation and controller complexity of successful and unsuccessful \PWM agents, depicted in the two bottom rows of   Figure \ref{fig:11}.
The successful agents have a much higher Morphological Computation over all. The Morphological Computation measures how much the next sensor states depend on the last sensor states, given the actuator nodes, and is calculated using the ideal world models. This means that the successful agents found strategies to move in their environment and use the interaction with the environment in a way, such that the next point in time is more predictable, more closely depends on the last sensory state, compared to the unsuccessful agents.

%\begin{figure}[htp]
%\includegraphics[width = 0.475\textwidth]{Figures/Figure12a.eps}
%\includegraphics[width = 0.475\textwidth]{Figures/Figure12b.eps}
%\vspace{-0.25cm}
%\caption{{The Morphological Computation and Integrated Information of the \PWM agents in case of successful
%agents on the left and unsuccessful agents on the right.}}
%\label{fig:12}
%\end{figure}

Additionally, the Integrated Information is overall higher in the case of the unsuccessful agents. There the agents have a lower Morphological Computation and we again observe an inverse correlation
between these two quantities. Previously we noted that the Integrated Information is not influenced by
the em-algorithm after the first steps, however, the observation made here refers to the value that
the algorithm reaches exactly during these first steps.

To conclude, if we have an \PWM agent with access to its correct world model and with a morphology which is well-adapted to its environment, then the \PWM agent has no need for a complex control architecture, a brain.

In order to further examine this connection between the quality of the world model and the need for a complex controller we additionally analyze agents that are only able to sample their ideal world models for a part of the total 20$\,$000 steps. These agents
sample the ideal world model and learn their behavior only up to a certain
point. After that point the world model stays fixed and the agents have to use this, possibly inaccurate,
world model to find the best behavior for the remainder of the 20$\,$000 steps.
We distinguish between 9 different cases, namely agents that sample the world model for 50, 100,
200, 500, 1000, 2000, 5000, 10000 or the full 20000 steps. 

\begin{figure}[htp]
\centering
\includegraphics[width = 0.975\textwidth]{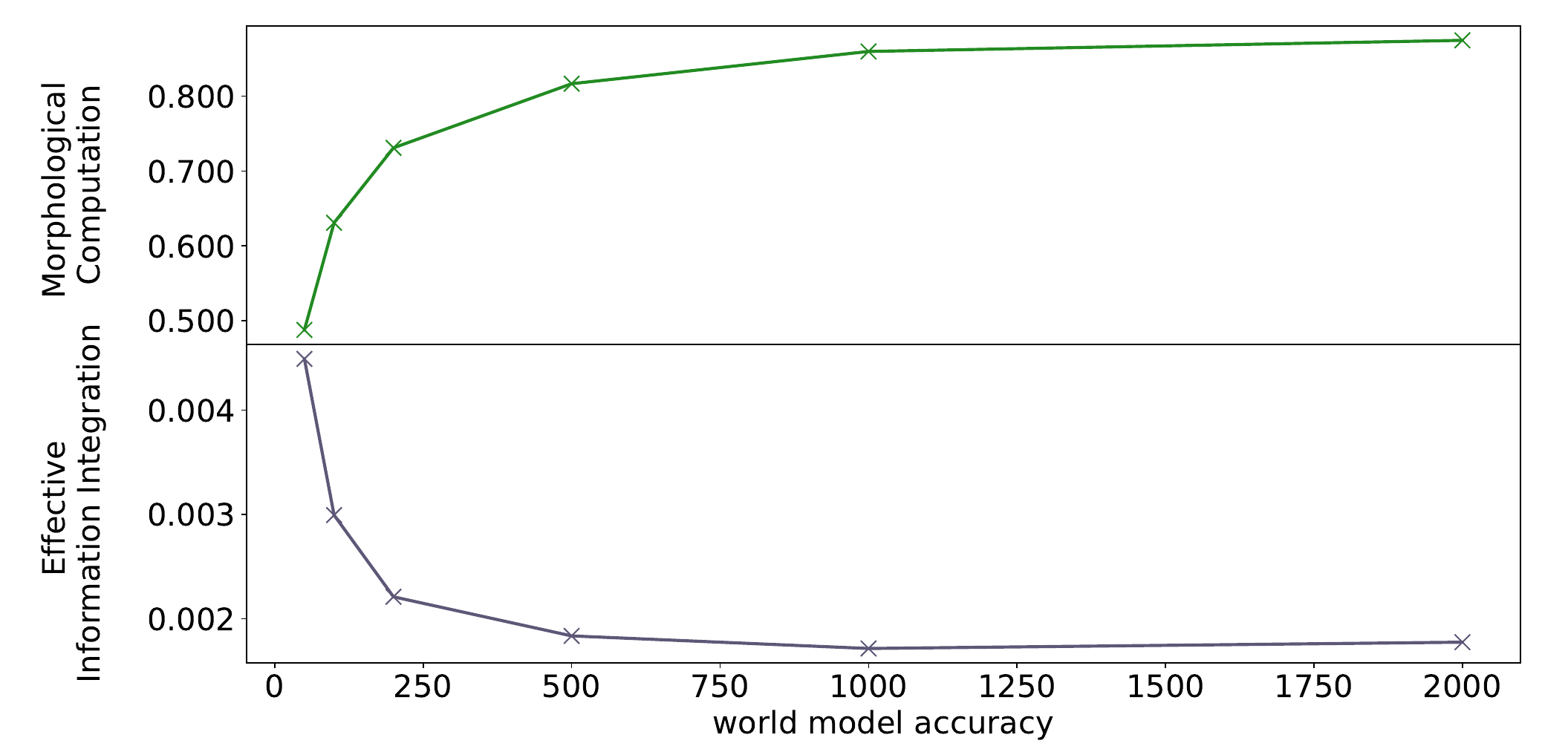}
\caption{{ Morphological Computation and effective information integration for the successful \PWM agents with a
varying accuracy of the world model.}}
\label{fig:13}
\end{figure}

In Figure \ref{fig:13} we highlight the relationship
between Morphological Computation and effective information integration, defined in equation \eqref{phi_eii}, with respect to the
accuracy of the world model on the x-axis. There we display the arithmetic mean over the different sensor length after 20$\,$000 steps.
While the Morphological Computation increases with the accuracy of the world model,
the effective information integration decreases.

\newpage

In the introduction we motivate
the intuition behind these concepts using the
example of a child learning to ride a bike. The
better the child understands the dynamics of
its environment, the more it can make use of
them and the faster it drives to stabilize the bike.
Hence, a better world model leads to a
higher Morphological Computation which then reduces the necessity for a complex controller.

This concludes the analysis of the accuracy of the ideal world model in relationship to the
information flow inside the agents. In the next section we discuss the \LWM  agents that 
\textcolor{changed}{additionally have to learn their internal world models}.
%PLOS does not support heading levels beyond the 3rd (no 4th level headings).

\subsection*{The Internal-World-Model  Agents} \label{Subsect:natural}
\begin{figure}[ht!]
\centering
\includegraphics[width = \textwidth]{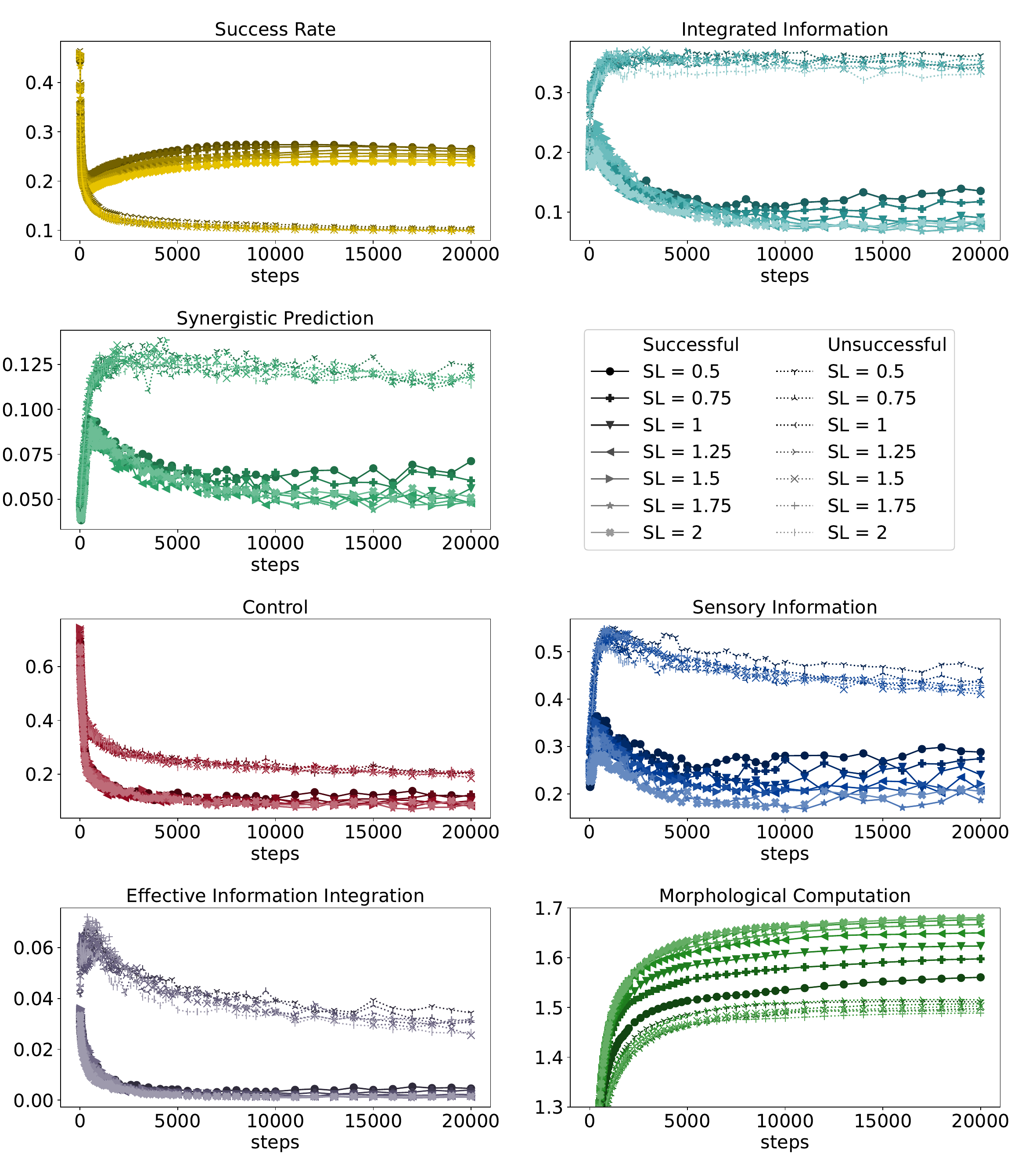} 
\caption{{The success rate, Integrated Information, synergistic prediction, control, sensory information, effective information integration and Morphological Computation
for the successful and unsuccessful \LWM agents.}}
\label{fig:14}
\end{figure}

Here we discuss the results for the \LWM  agents, which have to learn the dynamics of the world that are relevant to the agent. We first focus on the measures for the controller complexity, called Integrated Information and synergistic prediction. \textcolor{changed}{The Figure \ref{fig:14} depicts the success rate and all the information theoretic measures for the successful and unsuccessful \LWM agents}. Each line in these figures corresponds to a sensor length (SL) and is depicted with respect to the number of steps.
The unsuccessful agents have an Integrated Information value around 0.3 and 0.38 and a synergistic prediction around 0.1 and 0.14.

Now we compare these results to the Integrated Information value of the successful agents and we can observe that  there is a first increase in the Integrated Information and synergistic prediction values in the first 400 steps and then a strong decrease. After 20$\,$000 steps the Integrated Information value is roughly between 0.05 to 0.15 and the synergistic prediction lies between 0.04 and 0.08. Hence, in the case of the successful agents the complexity of the controller reduces to a much lower value over time, compared to the unsuccessful agents.

Following the observations of the previous section leads to the conclusion that a high controller complexity might be important as long as the agents have not been able to learn the correct world model. Without a correct world model the agents are not able to find a strategy that would allow them to optimally use their interaction with the environment.

In order to interpret these results in relation to the learning behavior of the agents further, we now
discuss the values for the sensory information, control and Morphological Computation.
These first two give insights to the effect the Integrated Information has on the action of the agent and combined lead to the effective information integration. \textcolor{changed}{In the two bottom rows of Figure \ref{fig:14}} we depict these four measures. 
The sensory information and control decrease with the number of steps taken for the successful, as well as the unsuccessful agents. However, there is a clear difference in the overall values of these measures, which leads to the effective information integration of the successful agents being around 0.002, while this value reaches in average 0.03 in case of the unsuccessful agents. Hence, the Integrated Information is not only higher for the unsuccessful agents, it also has more impact on the behavior of the agents.  

%\begin{figure}[h]
%\includegraphics[width = \textwidth]{Figures/Figure15c.eps}
%\includegraphics[width = \textwidth]{Figures/Figure15d.eps}
%\vspace{-0.8cm}
%\caption{{The measures for sensory information, control, effective information integration and Morphological
%Computation for the successful, \LWM agents in the top and for the unsuccessful agents in the bottom row.}}
%\label{fig:15}
%\end{figure}

\textcolor{changed}{In case of the Morphological Computation} we observe that the successful agents reach a higher Morphological Computation value, in average 1.64, compared to a value of 1.5 in case of the unsuccessful agents.  

These results support our hypothesis. A high controller complexity value seems to be important as
long as the agents have not been able to learn to interact with their environment. Hence, the Morphological Computation is lower for the unsuccessful agents while the complexity and involvement of the controller is higher. Now the question remains, whether a high controller complexity is really necessary for learning or just a byproduct of the Morphological Computation being low.
In order to clarify that point we now look at the split \LWM agents, which have a simplified control architecture.

\subsection*{Comparing the split and complete \LWM Agents}\label{Subsect:Limited}

The architecture of the split \LWM agents is depicted on the bottom right in Figure \ref{fig:3}. These agents are not able to integrate information between their controller nodes, hence the complexity of the controller solely depends on the structure of the internal world model. 

\newpage
We divide these agents into successful and unsuccessful ones by applying the criterion for success from the \LWM  agents, of $16.8\%$. The split agents perform worse and this does not lead to a $1/3, 2/3$ split. However, thereby we are able to directly compare complete \LWM  and split agents with a similar success rate.  

First, we consider the average success rates of the split and complete \LWM  agents. In addition, we compare them with the average success rate of agents that perform only random movements and do not learn at all. The results are given in Table \ref{table1}

\begin{table}[htp]
%\begin{adjustwidth}{-2.25in}{0in} % Comment out/remove adjustwidth environment if table fits in text column.
\centering
\begin{tabular}{l c c c }
& random  &   complete & split  \\
& movement & \textcolor{white}{..} \LWM \textcolor{white}{..} &\textcolor{white}{..}  \LWM \textcolor{white}{..} \\
& & agents & agents \\
average \\
 success rate & $\approx 7.95 \%$ & $\approx 15.21 \%$ & $\approx 8.01 \%$  \\ 
%success rate ideal agents &  & $\approx 32 \%$ & $\approx 30 \%$ 
\end{tabular}
\caption{
{\small Arithmetic mean of the SRs of agents with random movement, the complete and split \LWM agents.}} \label{table1}
%\end{adjustwidth}
\end{table}

The split \LWM agents perform in average barely better than the agents that move randomly. Note, that there is also a considerable difference in the number of  successful agents. Only approx. $2.1\%$ of split \LWM agents are successful, compared to $33.3\%$ of the complete ones.

In summary, the split agents perform only marginally better than agents that move purely at random and only very few split agents are successful.
This strongly supports the hypothesis that an increased controller complexity is necessary for learning.

\begin{figure}[htp]
\centering
\includegraphics[width = \textwidth]{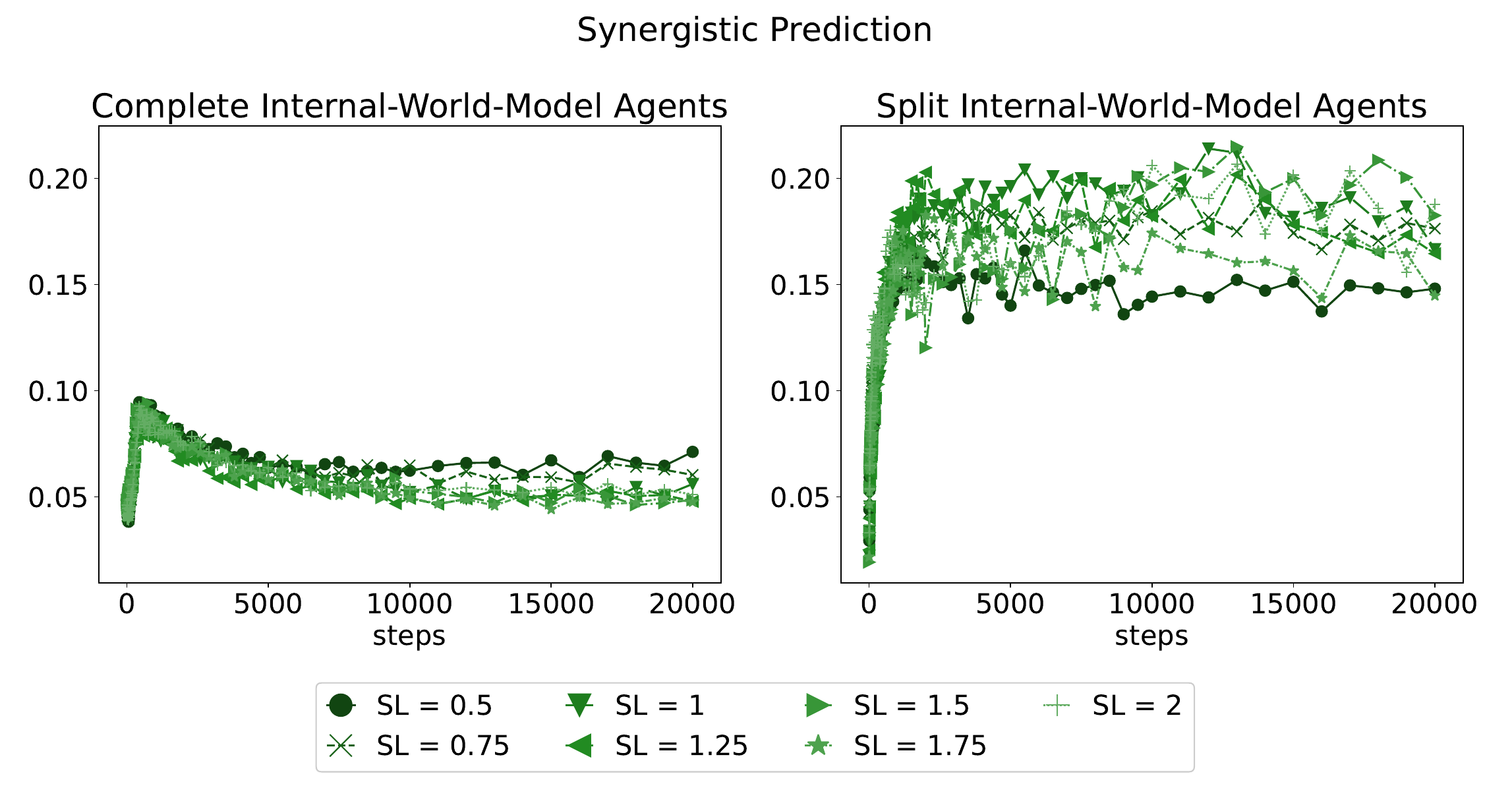}
\vspace{-0.25cm}
\caption{{Comparison of the synergistic prediction in case of the successful \LWM agents in the first and for the
successful split \LWM agents in the second column.}}
\label{fig:16}
\end{figure}

Additionally, we now focus on the internal world model of the few successful agents. Here, we compare the synergistic prediction of the successful, complete  \LWM  and the successful, split \LWM agents. The results are shown in Figure \ref{fig:16}.

The synergistic prediction quantifies how important the interaction of both influences, from the actuators as well as from the controller nodes, are for the prediction. It is noticeable that the synergistic prediction is much higher for the successful agents that are not able to integrate information. This leads to the conclusion that for these split agents the internal world model, and therefore the prediction process, has to combine the information from different sources and becomes much more complex. The complete \LWM  agents are able to integrate the information directly between their controller nodes and do not need such a complicated world model to have a complex controller.

\newpage
\section*{Conclusion} \label{Sect:Conclusion}

In this article we discuss the dynamics of the Morphological Computation and controller complexity in learning, embodied artificial agents. These agents move inside a racetrack and learn to not touch the walls. Using this simplistic example, we are able to analyze the different information flows inside the agents and especially examine the process of predicting the next sensory state. As a training algorithm we use an adapted em-algorithm that alternates between optimizing the behavior to reach a goal and updating the internal world model. \textcolor{changed}{This algorithm fits naturally in our framework, since it is information geometric in nature. Additionally, its geometric interpretation highlights the interplay between the goals of optimizing the behavior and the world model. However, for future work we intend to analyze the influence that more biologically plausible leaning algorithms have on the information flows inside the system.} 

The results of our experiment regarding the controller complexity and Morphological Computation support our previous publication \citep{Langer2021}, \textcolor{changed}{since we observe} the inverse correlation between them. These previous results lead to the insinuation that agents with a highly adapted morphology might have no use for a complex control architecture. There are many possible ways to address this notion. It might be that our tasks are simply too easy to solve, so that an agent truly only needs Morphological Computation in order to be successful. 
Another possibility is given by the authors of \citep{Pfeifer2009}

\begin{quote}
``The more the specific environmental conditions are exploited - and the passive dynamic walker is an extreme case - the more the agent’s success will be contingent upon them. Thus, if we really want to achieve brain-like intelligence,
the brain (or the controller) must have the ability to quickly switch to different
kinds of exploitation schemes either neurally, or mechanically through morphological change.''
\end{quote}

Hence, the agents might have no need for a controller, because they are only faced with one single task, namely avoiding the walls of their environment. Furthermore, the nature of the task might be too simplistic and it might have to require a higher order of understanding of its surroundings, so that the agents truly need to process the information from the environment. Therefore we will develop this approach further in order to explore these possibilities and apply it to more involved settings.   

Despite the simplicity of our example, we were able to offer an additional solution to the posed problem. We theorize that learning to predict the environment results in a necessity for a complex controller. \textcolor{changed}{Ideal-world-model} agents, which do not have to learn to predict their environment, do not require a complex controller at all, not even to learn our task. However, when their ability to form an accurate world model is restricted the involvement of the control architecture increases.

The \LWM  agents, on the other hand, show a necessity for an increased controller complexity in general. The controller complexity of the successful agents is first high, while the agents learn their world model, and then it decreases. We argue that this decrease could result from a rise in Morphological Computation that is facilitated by the correct world model. This is supported by the results for Morphological Computation, which are higher in the case of the successful agents. Hence, the two quantities, the controller complexity and the Morphological Computation, influence each other. 

Comparing the complete \LWM  agents with the split ones, which have a simplified controller and are not able to integrate information, leads to the observation that the latter agents are not able to predict the next sensory state as well. The split \LWM agents perform in average only marginally better than completely randomly moving agents and there is only a very small percentage of successful split \LWM agents. Hence, learning requires an increased controller complexity.

Furthermore, the few successful, split \LWM agents have a more complex prediction process. This process itself combines the information from the controller and the actuator nodes in order to form a prediction of the next sensory state. This again supports the claim that an agent needs to integrate its available information in order to learn. In this case the complex process is not directly between the controller nodes, but inside the internal world model.

\section*{Acknowledgment}
The authors acknowledge funding by Deutsche Forschungsgemeinschaft Priority Programme “The Active
Self” (SPP 2134).

\printbibliography

\appendix
\newpage
\section{Learning the Strategy and the World Model } \label{S1_Appendix} %Simultaneously

The learning algorithm applied to the \LWM agents in our experiments works by adapting the em-algorithm to incorporate two different goals. The em-algorithm is a well-known information geometric algorithm that iteratively projects to two different sets of probability distributions and thereby reduces the KL-divergence between them  \citep{Amari1995, Amari2007}. 

{\color{changed}In this article an agent learns inside the racetrack. Hence, the realized states $s_{t-1}, a_{t-1}, c_{t-1}$ are known at each step $t$ and can be used. To that end we need the following definitions.

Let $P_{c_t}(C_{t+1} \vert S_{t+1})$ be the distribution of $C_{t+1}$ conditioned on $S_{t+1}$ and a fixed state $c_{t}$, meaning that
$P_{c_t}(c_{t+1} \vert s_{t+1}) := P(c_{t+1} \vert s_{t+1}, c_t) $ for all $ s_{t+1}, c_{t+1} \in \mathcal{S} \times \mathcal{C}. $

The \PWM agents are able to use the sampled world model for the prediction, whereas the \LWM agents make use of their internal world model to arrive at the internal prediction $S'$. Both types of agents can optimize the distributions
$P_{c_t}(c_{t+1} \vert  s_{t+1}),  P(a_{t+1} \vert s_{t+1} , c_{t+1} )$ with $s_{t+1}, a_{t+1}, c_{t+1} \in \mathcal{S} \times \mathcal{A} \times \mathcal{C}.$ In addition, the \LWM agents also learn their internal world model given by the distribution $P(S'_{t+2} \vert A_{t+1}, C_{t+1})$.
 }

\textcolor{changed}{Furthermore, we add gaussian noise to the distribution $P(A_{t+1} \vert S'_{t+1} , C_{t+1} )$, because the em-algorithm can not gain a positive value again once it reaches a point where for some action $a_{t+1}$ the equality $P(a_{t+1} \vert s'_{t+1} , c_{t+1} ) = 0$ holds.}

\textcolor{changed}{Now we introduce the modified em-algorithm}. First we define one set for optimizing with respect to the goal. Let $S'^3$ be the variable indicating whether the agent is touching a wall. Then $s^{3} =1$ signifies that the agent is not touching a wall. Now let $X'_{t+1} = (S'_{t+1}, A_{t+1}, C_{t+1})$, then the goal manifold consists of all those probability distributions for which it is certain that the agent will not touch a wall at point $t+2$

\begin{equation}
\mathcal{M}_{G}^{P} (x_t) = \left\lbrace Q \in \mathcal{P}(\mathcal{X} \times \mathcal{S}) \vert Q(s^{3}_{t+2} =1 ) = 1  \right\rbrace.
\end{equation}

The second set consists of all the distributions that factor according to the agents, meaning that they consist of all the possible agents, given the current world model

\begin{equation}
\begin{split}
\mathcal{M}_{A}^{P} (x_t, \overline{P}) = \left\lbrace P \in \mathcal{P}^{\circ}(\mathcal{X} \times \mathcal{S}) \vert P(x_{t+1}, s_{t+2}) =  \overline{P}_{a_t, c_t}(s_{t+1}) \prod\limits_j P_{c_t}(c_{t+1}^j \vert s_{t+1}) \right. & \\ 
  \left.  \prod\limits_i P(a_{t+1}^i \vert s_{t+1}, c_{t+1}) \overline{P}(s_{t+2} \vert a_{t+1}, c_{t+1}), \, \forall (x_{t+1}, s_{t+2}) \in \mathcal{X} \times \mathcal{S}  \right\rbrace&,
  \end{split}
\end{equation}

where $\mathcal{P}^{\circ}$ is the interior of $\mathcal{P}$ and $\bar{P}$ indicates that this distribution is fixed.

In \citep{Langer2021} we iteratively project between these two sets in order to find the distribution in $\mathcal{M}_{A}^{P}$ that is closest to $\mathcal{M}_{G}^{P} $. This would be the distribution that describes a valid agent and has a high likelihood of achieving the goal. This approach is also called planning as inference, \cite{Attias, Toussaint2006, Toussaint2009}. The approach is guaranteed to converge, but might converge to a local minimum. 

In our case we want to adapt this approach in order to simultaneously learn the internal world model. The distribution $P(S'_{t+1} \vert A_t, C_t)$ predicts the next sensory input and reflects therefore the agent's understanding of its environment. Hence we want to optimize our world model such that

\begin{equation}
P(S'_{t+2} \vert  S_{t+1}, A_{t+1} ) =  \tilde{P}(S_{t+2} \vert S_{t+1}, A_{t+1}),
\end{equation}

where $\tilde{P}$ is the sampled, ideal world model. \textcolor{changed}{We sample the distributions $P(S_t, A_t, C_t)$ and $P(S_{t+1} \vert S_t, A_t)$ as described in more detail by \citet{Ghazi-Zahedi2010} and we mark sampled distributions by a tilde $\tilde{P}$.}
Note, that we require the goal to be defined by a joint distribution, not a conditional, hence the actual optimization works with 

\begin{equation}
\overline{P}(S_{t+1}, A_{t+1}) \otimes P(S'_{t+2} \vert  S_{t+1}, A_{t+1} ) = \overline{P}(S_{t+1}, A_{t+1}) \otimes \tilde{P}(S_{t+2} \vert S_{t+1}, A_{t+1}).
\end{equation}

The joint distribution $\overline{P}(S_{t+1}, A_{t+1})$ is fixed to the joint distribution resulting from the last step in the algorithm. 

The conditional distribution $P(S'_{t+2} \vert  S_{t+1}, A_{t+1} )$ can be calculated as 

\begin{equation}
P(s_{t+2} \vert  s_{t+1}, a_{t+1} ) =\dfrac{ \sum\limits_{c_{t+1}} P_{a_t, c_t}(s_{t+1}) P_{c_t}(c_{t+1} \vert s_{t+1}) P(a_{t+1} \vert s_{t+1}, c_{t+1} ) P(s_{t+2} \vert c_{t+1}, a_{t+1}) }{\sum\limits_{c_{t+1}} P_{a_t, c_t}(s_{t+1}) P_{c_t}(c_{t+1} \vert s_{t+1}) P(a_{t+1} \vert s_{t+1}, c_{t+1} )}.
\end{equation}

Then the world goal manifold results in

\begin{equation}
\begin{split}
\mathcal{M}_{G}^{W} (x_t, \overline{P}) = \left\lbrace Q \in \mathcal{P}(\mathcal{X} \times \mathcal{S}) \vert Q(s_{t+2}, s_{t+1}, a_{t+1} )  = \overline{P}(s_{t+1}, a_{t+1})\tilde{P}(s_{t+2} \vert s_{t+1}, a_{t+1}), \right. \\ \left.  \, \forall (s_{t+1}, a_{t+1}, s_{t+2}) \in \mathcal{S} \times \mathcal{A} \times \mathcal{S}  \right\rbrace.
\end{split}
\end{equation}

Similar to the agent manifold above, we also define a world agent manifold

\begin{equation}
\begin{split}
\mathcal{M}_{A}^{W} (x_t, \overline{P}) = \left\lbrace P \in \mathcal{P}^{\circ}(\mathcal{X} \times \mathcal{S}) \vert P(x_{t+1}, s_{t+2}) = P_{a_t, c_t}(s_{t+1}) \prod\limits_j \overline{P}_{c_t}(c_{t+1}^j \vert s_{t+1}) \right. \\
\left. \prod\limits_i \overline{P}(a_{t+1}^i \vert s_{t+1}, c_{t+1}) P(s_{t+2} \vert a_{t+1}, c_{t+1}), \, \forall (x_{t+1}, s_{t+2}) \in \mathcal{X} \times \mathcal{S}  \right\rbrace.
\end{split}
\end{equation}

Note that we can define a full agent manifold by 

\begin{equation}
\begin{split}
\mathcal{M}_{A} (x_t) = \left\lbrace P \in \mathcal{P}^{\circ}(\mathcal{X} \times \mathcal{S}) \vert P(x_{t+1}, s_{t+2}) = P_{a_t, c_t}(s_{t+1}) \prod\limits_j P_{c_t}(c_{t+1}^j \vert s_{t+1}) \right. \\
\left. \prod\limits_i P(a_{t+1}^i \vert s_{t+1}, c_{t+1}) P(s_{t+2} \vert a_{t+1}, c_{t+1}),  \, \forall (x_{t+1}, s_{t+2}) \in \mathcal{X} \times \mathcal{S} \right\rbrace
\end{split}
\end{equation}

and then $\mathcal{M}_{A}^{W}  \subset \mathcal{M}_{A}$ and $\mathcal{M}_{A}^{G} \subset \mathcal{M}_{A} $.  Similarly, we can also define a full world goal manifold

\begin{equation}
\begin{split}
\mathcal{M}_{G}^{W} (x_t) = \left\lbrace Q \in \mathcal{P}(\mathcal{X} \times \mathcal{S}) \vert \exists P \in \mathcal{P}(\mathcal{X}) \text{ : } Q(s_{t+2}, s_{t+1}, a_{t+1} )  = P(s_{t+1}, a_{t+1}) \right. \\ 
\left. \tilde{P}(s_{t+2} \vert s_{t+1}, a_{t+1})   \, \forall (s_{t+1}, a_{t+1}, s_{t+2}) \in \mathcal{S} \times \mathcal{A} \times \mathcal{S}  \right\rbrace.
\end{split}
\end{equation}

\begin{figure}[ht]
\centering
\includegraphics[width = 0.75\textwidth]{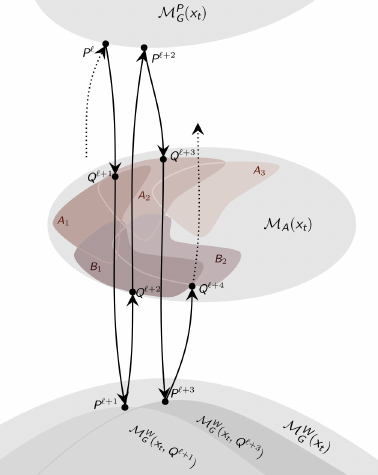}
\caption{{ The modified em-algorithm for optimizing the behavior and the internal world model simultaneously. } Here $A_1 = \mathcal{M}_{A}^{G} (x_t, P^{l})$, $ A_2 = \mathcal{M}_{A}^{G} (x_t, P^{l+2})$, $ A_3 = \mathcal{M}_{A}^{W} (x_t, P^{l+4})$, $ B_1 = \mathcal{M}_{A}^{W} (x_t, P^{l+1})$, $ B_2 = \mathcal{M}_{A}^{W} (x_t, P^{l+3})$. }
\label{fig:17}
\end{figure}

Now we are able to define the algorithm depicted in Figure \ref{fig:17}.

The first step is to project to $\mathcal{M}_G^P(x_t)$ via an $e$-projection

\begin{equation}
Q^0 = \arginf\limits_{Q \in \mathcal{M}_G^P(x_t)} D(Q \parallel P^0).
\end{equation}

Then we project with an $m$-projection to $\mathcal{M}_{A}^{P} (x_t, P^0) $

\begin{equation}
P^1 = \arginf\limits_{P \in \mathcal{M}_{A}^{P} (x_t, P^0)} D(Q^0 \parallel P).
\end{equation}

Up to this point, this is the standard em-algorithm. Now instead of projecting to $\mathcal{M}_G^P(x_t)$ again, we update the internal world model by projecting to $\mathcal{M}_G^W(x_t, P^1)$ with an $e$-projection

\begin{equation}
Q^1 = \arginf\limits_{Q \in \mathcal{M}_W^P(x_t, P^1)} D(Q \parallel P^1).
\end{equation}

Afterwards we project, with an $m$-projection to $\mathcal{M}_{A}^{W} (x_t, P^1)$

\begin{equation}
P^2 = \arginf\limits_{P \in \mathcal{M}_{A}^{W} (x_t, P^1)} D(Q^1 \parallel P).
\end{equation}

Then we start the process over by projecting to $\mathcal{M}_{G}^{W} (x_t)$ as depicted in Figure \ref{fig:17}. 

Note that this algorithm is not guaranteed to converge. Since we are interested in agents that learn while performing a task, we perform only of the 5 optimization steps, as described above, after each step of the agent. Therefore a convergence is not needed in our scenario. 
\end{document}